\documentclass[aps,prb,preprint,onecolumn,superscriptaddress]{revtex4}
\usepackage{graphics}
\usepackage{epsfig}
\usepackage{units}
\usepackage{color}
\usepackage{amsmath}

\begin{document}

\title{Electron-phonon interaction in the dynamics of trap-filling in quantum dots.}

\author{R. Carmina Monreal}
\email[Corresponding Author: ]{r.c.monreal@uam.es}
\affiliation{Departamento de F\'{\i}sica Te\'{o}rica de la Materia Condensada and Condensed Matter Physics Center (IFIMAC), Universidad Aut\'{o}noma de Madrid, E-28049 Madrid, Spain}



\begin{abstract}
We analyze theoretically the effects of electron-phonon interaction in the dynamics of an electron that can be trapped to a localized state and detrapped to an extended band state of a small quantum dot (QD) using a simple model system. It consists of a one-dimensional tight binding linear chain of a few sites, having a discrete set of energy levels mimicking the discrete levels of the conduction band of the QD, that is connected at its end to another site, the trap, having a single energy level well below the conduction band, where the electron is allowed to interact with a local phonon of a single frequency. 
In spite of its simplicity the time dependent model has no analytical solution but a numerically exact one can be found producing a rich dynamics. The electronic motion is quasi-periodic in time, with oscillations around a mean value that are basic characteristics of the weak and strong coupling regimes of electron-phonon interaction and set the time scales of the system. Using values of the parameters appropriate for defects in semiconductor QDs, we find these time scales to range typically from tenths of picoseconds to a few picoseconds. 
The values of the time averaged trap occupancy strongly depend on the the strength of the electron-phonon interaction and can be as large as 40$\%$ when the coupling is most efficient, independently of other parameters. 
An interesting result of the present work is the formation of resonances at specific values of the electron-phonon coupling parameter that only exist when several levels are allowed to coherently cooperate in the filling of the trap. They are characterized by a trap occupancy that is a periodic function of time with large amplitude and period picturing an electron that is periodically trapped and detrapped. 
We conclude that the formation of these resonances is a robust consequence of electron-phonon interaction in small systems.  Electron-phonon interaction is an efficient mechanism that can provide ca. 50$\%$ filling of a deep trap state on a subpicoseconds to picoseconds time scale, much faster than radiative decay occurring in time scales of tens of picoseconds to nanoseconds, while the occupancy of this state will be smaller than ca. 1$\%$ in the absence of electron-phonon coupling. 
\end{abstract}

%

\maketitle
\section{Introduction}

Optical properties of semiconductor nanocrystals (quantum dots QDs)\cite{Pichaardy, LEED3} are attracting an increasing interest for applications in optoelectronic devices\cite{Opto1,Opto2}, solar cells \cite{Solar1} and light emitting diodes \cite{LEED1,LEED2} among others.
It is experimentally well established that a colloidal QD is formed by a nucleus with the perfect lattice symmetry of the bulk semiconductor material and a surface.  However, the band structure of a QD of a few nanometers in size is discrete due to quantum-size effects. A consequence of quantum confinement of electrons is the phonon bottleneck. Excited electrons are long lived excitations in a small QD because they would couple very inefficiently to phonons due to the large mismatch between the inter-level energy spacing and the optical phonon energy \cite{Nozik_PhysE}. 
Light emission in colloidal QDs is conditioned by the presence of defects and/or impurities at the surface. 
In these small systems, the large surface to volume ratio results in relative abundance of surface defects of varied microscopic nature, such as surface defects, impurity atoms or surface ligands, generically called "traps", that produce states in the bandgap \cite{LEED3, JPCB_104_1715_vanDijken}.
Once the QD is excited by light across the bandgap, these localized states can trap charge carriers and alter the subsequent light emission. One paradigmatic case is 
ZnO for which its luminescent properties are known to depend upon the atmosphere surrounding the particles
\cite{ JPCB_104_1715_vanDijken, ChemPhysLett_122_507_koch, JPhysChem_91_3789_bahnemann, JLumin_90_123_vanDijken, JPCB_104_4355_vanDijken}, 
the charge on them \cite{JPCC_114_220_stroyuk, JPCC_115_21635_yamamoto,
JPCC_116_20633_cohn}, the passivation of surface defects \cite{JPCB_109_20810_norberg} or chemical modifications at the surface \cite{JACS_123_11651_shim}.
A very illustrative picture of a small colloidal QD and its band structure is shown in Figure 1 of Ref. \cite{AccChemRes_49_2127_kilin}.  

The atomistic nature of the trap states has been investigated in many cases. A number of intrinsic point defects and impurities have been proposed and theoretically 
studied  by Janotti and Van der Walle in the bulk of ZnO \cite{Janotti_RPP,Janotti_JAP} and of other semiconductors as well \cite{Janotti_PRB} using first-principles Density Functional Theory. In the case of QDs, the theory is focused on calculations of the electronic band structure of clusters with sizes smaller than ca. 3 nm in diameter with and without ligands at the surface, and at different levels of 
sophistication (we refer the reader to references \cite{AccChemRes_49_2127_kilin, JPCL_8_5209_giansante, JPCL_Peyton} and references therein)
but such calculations cannot be extended to larger systems nowadays. 
The existence of local lattice relaxations at the trap site immediately bears to mind electron-phonon interaction. Indeed the vibronic interaction has already been included in calculations of photoluminescence line shapes of a number of defects in bulk semiconductors \cite{Janotti_PRL, Janotti_APL}. All of these studies have in common that they focus on the static electronic or electronic plus vibronic configurations of the system.

However, a key point to understand the light emission properties of QDs is to know the time scales and efficiencies of possible mechanisms competing with radiative decay in the filling of the trap because if these mechanisms were much faster the photoluminescence will be quenched \cite{JPCC_116_20633_cohn}.  
A somehow related phenomenon is the fluorescence intermittency (known as blinking), observed in many classes of QDs, in which there are periods of time where the QD is "dark" in spite of continuous illumination while it is "bright" in other periods.
Transient states \cite{PRL_Cook} and tunneling between "dark" and "bright" states \cite{PRL_Efros, PRB_Kuno} have been proposed as mechanisms originating blinking.
In a recent calculation by du Foss\'e at al. \cite{ChemMat_duFosse} the initial photoexcitation of an electron to the conduction band of a CdSe QD of ca. 2 nm in diameter leads to a transient trap state localized in a Cd-Cd dimer that appears and disappears on the picoseconds time scale. 
In another related context, the analysis of the electronic transport through a molecular junction where electrons are coupled to phonons, some approximations have been developed to convert the full time dependent problem into a numerically tractable one (see Refs. \cite{PRB_Maier, PRB_Ferdinand, PRB_Ruben} and references therein). Time dependent theoretical investigations are scarce due to the enormous complexity of the problem.

The aim of the present work is to analyze the effects of electron-phonon interaction in the dynamics of an electron that can be trapped to a localized state and detrapped to an extended band state using a simple model, that will allow us to find general characteristics of its time evolution valid for different material systems.
The physical system we have in mind is a small colloidal QD as described above. In our model, the core of the QD is a tight binding linear chain of a few identical sites whose discrete set of eigenenergies mimics the discrete levels of the conduction band of a small QD.  
Substitution of a real QD by a few discrete energy levels has been frequent in the literature, in particular in the analysis of the phonon bottleneck problem
\cite{Inoshita_PRB, Kral_PRB,Stauber_PRB, Vasilevskiy_PRB}.
Since traps are frequently located at the surface of QDs, the trap will be modeled by another site at the end of the chain, having a single energy level well below the conduction band, as is the case for deep traps in semiconductor QDs
\cite{JPCB_104_1715_vanDijken, JLumin_90_123_vanDijken, JPCB_104_4355_vanDijken, JPCB_109_20810_norberg,Janotti_PRL,Janotti_APL,ChemMat_duFosse}. 
An electron at the trap site interacts with a local phonon of a single frequency while electron-phonon interaction within the chain is neglected because of the phonon bottleneck effect.
The use of  this one-dimensional model is justified by the detailed calculations of 
 Refs. \cite{Janotti_PRL, Janotti_APL}, where it is shown  that the multidimensional vibrational problem of a defect in a lattice with many degrees of freedom can be conveniently mapped onto an effective one-dimensional problem including one or two phonon modes. 
Moreover in this work we consider the simplest possible case of having a single electron in the system so electron spin is irrelevant.  This can be the case when the QD is illuminated with light of very low intensity so basically only one electron is excited to the conduction band or if the QD is charged with a very low density of electrons or if the system is in the presence of a strong magnetic field in which the physics occurs in energy scales much smaller that the Zeeman splitting energy. 
In the photoexcitation of a QD an electron is excited to the conduction band. We take this event as our starting point and consider an electron in one of our energy levels at time t=0. The electron is subsequently allowed to hop back and forth to the trap and we calculate numerically the evolution of the coupled electron-phonon system as a function of time.  The physical parameters of our problem which are the first neighbors hopping parameter within the linear chain, $\gamma$, the hopping between the last site of the chain and the trap site, $V_0$, the energy level of the trap site, $\epsilon_T$, the phonon frequency, $\omega_0$, and the strength of the electron-phonon interaction, $\lambda$, are in principle unrestricted but we will use values appropriate for defects in semiconductors QDs \cite{Janotti_PRL, Janotti_APL}.
This minimal model allows us to investigate analytical and numerically both the short and the long time dynamics of the electron, which will be presently infeasible for first-principles type of Hamiltonians.
The paper is organized as follows. In Section \ref{sec:theory} we expound the theory. We start with the simplest possible system consisting of two sites, the dot and the trap sites. In spite of its simplicity, this problem has no exact analytical solution but it can be approximated in different limits which allows us to pinpoint the physical parameters controlling the time evolution of the coupled electron-phonon system that set the time scales of the problem. 
We then consider a linear chain of three sites and, consequently, three energy bands. 
Section \ref{sec:results} is devoted to the presentation of the results of our numerical calculations. Starting with the two-sites system, we find a trap occupancy that is  a quasi-periodic function of time and this motion is set on a time scale of 0.1-4 picoseconds. The values of the time averaged trap occupancy depend on $\lambda/\omega_0$ and can be as large as 60$\%$ when the electron-phonon coupling is most efficient. 
The main characteristics of the electronic motion found for the two level system remain for the four-sites system, as well. One of the most interesting result of the present work is the formation of resonances. They are characterized by a trap occupancy that is a periodic function of time with large amplitude and period, that occur at specific values of $\lambda$ but only for $V_0 \geq \omega_0$, when several levels are allowed to coherently cooperate in the filling of the trap.
We have found them in both the two-sites and four-sites systems and for any value of the trap energy level.  We conclude that the formation of these resonances is a robust consequence of electron-phonon interaction in small systems.  Electron-phonon interaction is an efficient mechanism that can provide ca. 50$\%$ filling of a deep trap state on a picoseconds or subpicoseconds time scale, much faster than radiative exciton decay occurring in time scales of $\simeq 20$ ps \cite{JPCC_116_20633_cohn} to $\simeq 10$ ns \cite{Efros_NatNano}, while the occupancy of this state will be smaller than ca. 1$\%$ in the absence of electron-phonon coupling. 

Natural units $\hbar=e=m_e=1$ are used except otherwise indicated.

\section{Theory}
\label{sec:theory}
\subsection{Two-sites system}

In the most simple case the quantum dot is modeled by a single site having one energy level $\epsilon_D$, while the trap site
has one energy level $\epsilon_T$ coupled to phonons of energy $\omega_0$. The electron can hop from site to site with hopping parameter $V_0$. The Hamiltonian is the spinless Holstein Hamiltonian \cite{Holstein} and reads 

\begin{equation}
\hat H=\hat H_0+ \hat H_{hop},
\label{Ham-1}
\end{equation}
where the first term on the right hand side of Eq. (\ref{Ham-1}) is the Hamiltonian of the uncoupled dot and trap subsystems  

\begin{equation}
\hat H_0=\epsilon_D \hat n_D+ \hat H_T,
\label{Ham-0}
\end{equation}
and where the trap is modeled by the following Hamiltonian

\begin{equation}
\hat H_T =\epsilon_T \hat n_T+ \omega_0 \hat b^{\dagger} \hat b+ \lambda \hat n_T(\hat b^{\dagger}+\hat b),
\label{Ham-T}
\end{equation}
which describes the electron at the local trap site coupled with phonons, with a coupling constant $\lambda$. 
The last term in Eq. (\ref{Ham-1}) describes the electron hopping between the trap and the dot 

\begin{equation} 
\hat H_{hop}= V_0 (\hat c_{D}^{\dagger} \hat c_{T} +\hat c_{T}^{\dagger} \hat c_{D}). 
\label{H-int2}
\end{equation}

In Eqs. (\ref{Ham-T}) and (\ref{H-int2}) $\hat c_{D}$ ($\hat c_{D}^{\dagger}$) and $\hat c_{T}$ ($\hat c_{T}^{\dagger}$)
are the annihilation (creation) operators for electrons at the dot and trap respectively, $\hat n_D=\hat c_{D}^{\dagger} \hat c_{D}$ and 
$\hat n_T=\hat c_{T}^{\dagger} \hat c_{T}$ being their respective number operators,
$\hat b$ ($\hat b^{\dagger}$) are the annihilation (creator) operators for phonons, $\hat b^{\dagger} \hat b$ being their number operator. 
The time-dependent Sch\"ordinger equation for Hamiltonian Eq. (\ref{Ham-1}) will be solved for $t> 0$ with the initial condition that the electron is in the dot with zero phonons in the system at $t=0$.

Before analyzing the solution of the full time-dependent problem, it is useful to consider  
two limits that have an exact solution.
On the one hand, in the absence of electron-phonon coupling ($\lambda=0$), we have the text book problem of one electron hopping between levels $\epsilon_D$ and $\epsilon_T$.
If the electron starts in the dot at $t=0$ the occupancy of the trap oscillates in time according to Rabi's formula

\begin{equation}
n_R(t)=\frac{V_{0}^{2}}{V_{0}^{2}+(\frac{\epsilon_T-\epsilon_D}{2})^2} sin^2  \Big(t\sqrt{V_{0}^{2}+(\frac{\epsilon_T-\epsilon_D}{2})^2}\Big),
\label{nT-Rabi}
\end{equation}
with (Rabi) period

\begin{equation}
 T_R =\frac{\pi}{\sqrt{V_{0}^{2}+\left(\frac{\epsilon_T-\epsilon_D}{2}\right)^2}}
\label{T-Rabi}.
\end{equation}

On the other hand, in the absence of hopping ($V_0=0$) we have the so called atomic limit in which a single electron in a localized level interacts with phonons. 
The stationary solution of Eq. (\ref{Ham-T}) \cite{Mahan} yields the renormalization of the electron energy to $\tilde \epsilon_T=\epsilon_T-\frac{\lambda^2}{\omega_0}$ 
and a density of states which, at zero temperature, reads 

\begin{equation}
\rho_{T}(\omega)=\lim_ {\eta \rightarrow 0} Im e^{-g} \sum_{n=0}^{\infty}  \frac{g^n}{n!} \frac{1}{\omega-\tilde \epsilon_T-n \omega_0+i\eta}.
\end{equation}
where $g=(\frac{\lambda}{\omega_0})^2$ is the Huang-Rhys parameter and $Im$ stands for imaginary part.
The density of states consists of peaks at
$\tilde \epsilon_T+n \omega_0$, with weights distributed according to the Poisson distribution function. 
It has a maximum at the value of $n$ such as $n \approx g$ with a width $\approx 2g$. 
Hence our full problem deals with an electron that can hop between the dot level and a manyfold of levels at the trap site as a function of time. 

One case where an approximate solution of the full time-dependent problem can be found  is the quantum strong coupling regime
defined by $\lambda>>\omega_{0}$ and $\lambda>>V_{0}$. 
In this case it is
convenient to apply to Hamiltonian Eq. (\ref{Ham-1}) the
canonical transformation \cite{Lang_Firsov}
$\tilde{H}=\hat{S}\hat{H}\hat{S}^{-1}$ with $\hat{S}$ given by 

\begin{equation}
\hat{S}=exp[\frac{\lambda}{\omega_0}(\hat {b}^{\dagger}-\hat {b})\hat{n}_{T}],
\label{eq-S}
\end{equation}
which transforms electronic and bosonic operators as

\begin{eqnarray}
\tilde{c}_T & = & \hat{S}\hat{c}_T \hat{S}^{-1}=\hat{c}_T
exp[-\frac{\lambda}{\omega_0}(\hat {b}^{\dagger}-\hat {b})], 
\nonumber \\
\tilde{c}_D & = & \hat{S}\hat{c}_D \hat{S}^{-1}=\hat{c}_D,  
\nonumber \\
\tilde{b} & = & \hat{S} \hat b \hat{S}^{-1}=\hat {b}-\frac{\lambda}{\omega_0}
\hat{n}_{T} .  
\nonumber \\ 
\label{op-tilde}
\end{eqnarray}

Note that Eqs.(\ref{op-tilde}) imply that the number operators for
electrons in the dot and in the trap remain unchanged.
Then, the transformed Hamiltonian reads

\begin{equation}
\tilde H={\tilde H}_0+{\tilde H}_{hop},
\label{H-tilde}
\end{equation}
where

\begin{equation}
{\tilde H}_0= \epsilon_{D} \hat{n}_{D}+\tilde{\epsilon}_T \hat{n}_{T} +
\omega_0 {\hat b}^{\dagger}{\hat b},
\label{H0-tilde}
\end{equation}
with $\tilde\epsilon_T=\epsilon_T-\lambda^2/\omega_0$
representing the renormalization of the trap energy level due to
its coupling with the local phonons defined above
and

\begin{equation}
{\tilde H}_{hop}=V_{0}(\hat{c}_{D}^{\dagger} {\tilde c}_{T}+ {\tilde c}_{T}^{\dagger}\hat {c}_{D}).
\label{H-int-tilde}
\end{equation}

To solve the time-dependent problem we chose the orthonormal basis set of eigenstates of ${\tilde H}_0$

\begin{equation}
\{|\varphi_{D,n}\rangle, |\varphi_{T,n}\rangle \}_{n=0,1,2...},
\end{equation}
where $|\varphi_{D,n}\rangle $ and $|\varphi_{T,n}\rangle $ describe the electron at the dot site or at the trap site plus $n$ phonons, respectively, 
and are written in a second-quantization representation as 

\begin{eqnarray}
|\varphi_{D,n}\rangle &= &|1 0\rangle \otimes|n \rangle,  \nonumber \\
|\varphi_{T,n}\rangle &= &|0 1\rangle \otimes|n \rangle.  \nonumber \\ 
\label{basis}
\end{eqnarray}

The time-dependent wave function is then written as a linear combination of the basis states

\begin{equation}
|\tilde \psi(t) \rangle= \sum_{n=0}^{\infty}[\tilde a_{D,n}(t)|\varphi_{D,n}\rangle +\tilde a_{T,n}(t)|\varphi_{T,n}\rangle],
\label{psi-t}
\end{equation}
and the time-dependent Sch\"ordinger equation for Hamiltonian Eq. (\ref{H-tilde}) is projected onto the basis, leading to the following system of coupled linear differential equations

\begin{eqnarray}
\frac{d \tilde{a}_{D,n}(t)}{dt}&=&-i(\epsilon_D+n\omega_0)\tilde{a}_{D,n}(t)-i \sum_{m=0}^{\infty} W_{mn} \tilde{a}_{T,m}(t), \nonumber \\
\frac{d \tilde{a}_{T,n}(t)}{dt}&=&-i(\tilde{\epsilon}_T+n\omega_0) \tilde{a}_{T,n}(t)-i\sum_{m=0}^{\infty} W_{nm} \tilde{a}_{D,m}(t), 
\nonumber \\
\label{equ-t-tilde}
\end{eqnarray}
where the non-zero matrix elements of $\tilde{H}_{hop}$, $W_{nm}$,  are given by

\begin{eqnarray}
 W_{nm}& = &V_0 \langle \varphi_{T,n}|{\tilde c}_{T}^{\dagger} \hat{c}_{D}|\varphi_{D,m}\rangle=V_0 e^{-\frac{1}{2}g} \left(\frac{\lambda}{\omega_0}\right)^{n-m} \sum_{l=max (0, m-n )}^{m} (-g)^{l} \frac{\sqrt{n! m!}}{l! (m-l)! (l+n-m)!}, \nonumber \\
W_{mn}&=&V_0 \langle \varphi_{D,n}|\hat{c}_{D}^{\dagger} {\tilde{c}}_{T}|\varphi_{T,m}\rangle.   \nonumber \\
\label{Wnm}
\end{eqnarray}

For sufficiently large values of $g$, the matrix elements, being proportional to $exp(-\frac{g}{2})$, will be small so that we can perform standard time-dependent perturbation theory. The results depend on whether there exists a natural number $n_r$ such that the trap sublevel $\tilde{\epsilon}_T+n_r\omega_0$ is
exactly in resonance with the dot level $\epsilon_D$ ($\tilde\epsilon_T+n_r \omega_0-\epsilon_D=0$). If there is such sublevel the dominant coefficient 
in the wave function expansion is 
$\tilde{a}_{T,n_r}(t)$, yielding the trap occupancy as

\begin{equation}
n_{T}(t)\approx |\tilde{a}_{T,n_r}(t)|^2 = sin ^2[t V_0 e^{-\frac{g}{2}} (\frac{\lambda}{\omega_0})^{n_r}\frac{1}{\sqrt{n_{r}!}}].
\label{nT-res}
\end{equation}

Comparing Eq. (\ref{nT-res}) with the Rabi's formula of Eq. (\ref{nT-Rabi}) for $\epsilon_T-\epsilon_D=0$, we see that  the hopping to/from the dot to a resonant trap sublevel proceeds via a renormalized value of the hopping given by

\begin{equation}
\tilde{V}(n)= V_{0} \sqrt{ e^{-g} \frac{g^{n}}{n!}},
\label{V-tilde}
\end{equation}
with $n=n_r$. When none of the trap sublevels are in resonance with the dot level, we calculate the coefficients $\tilde{a}_{T,n}(t)$ to first order in $W_{nm}$ obtaining

\begin{equation}
n_{T}^{(1)}(t) = \sum_{n=0}^{\infty} |\tilde{a}_{T,n}^{(1)}(t)|^2 =
\sum_{n=0}^{\infty} \frac{\tilde{V}(n)^2}{\left(\frac{\tilde \epsilon_T+n \omega_0-\epsilon_D}{2}\right)^2}
sin^{2}[t \frac{\tilde \epsilon_T+n \omega_0-\epsilon_D}{2}],
\label{nT-nores}
\end{equation}
with $\tilde{V}(n)$ given by Eq. (\ref{V-tilde}). 
The comparison of Eqs.(\ref{nT-nores}) and (\ref{nT-Rabi}) strongly suggests that, for  $\tilde V(n) \ll \left(\frac{\tilde \epsilon_T+n \omega_0-\epsilon_D}{2}\right)$, the hopping to/from the dot to any of the trap sublevels also proceeds via the renormalized value  $\tilde{V}(n)$  with a characteristic period of oscillation
 
\begin{equation}
\tilde{T}(n)= \frac{\pi}{\sqrt{\tilde V^2(n)+\left(\frac{\tilde \epsilon_T+n \omega_0-\epsilon_D}{2}\right)^2}}.
\label{T-tilde}
\end{equation}

Eq. (\ref{T-tilde}) is Eq. (\ref{T-Rabi}) with $\epsilon_T$ substituted by $\tilde \epsilon_T+n \omega_0$ and $V_0$ substituted by $\tilde{V}$ and reproduces the period of the oscillations in Eqs.(\ref{nT-res}) and (\ref{nT-nores}) in the limits  $\tilde \epsilon_T+n_{r} \omega_0-\epsilon_D=0$ and  
$\tilde V(n) \ll \left(\frac{\tilde \epsilon_T+n \omega_0-\epsilon_D}{2}\right)$, respectively.

To solve the time-dependent problem for any value of the parameters, we chose the orthonormal basis set of Eqs. (\ref{basis}).  
The time-dependent wave function $|\psi(t) \rangle $ is written as a linear combination of the basis states, as in 
Eq. (\ref{psi-t}), with coefficients $a_{D,n}(t)$ and $a_{T,n}(t)$, 
and the time-dependent Sch\"ordinger equation for Hamiltonian Eqs.(\ref{Ham-1}-\ref{H-int2}) is projected onto the basis, leading to the following system of coupled linear differential equations

\begin{eqnarray}
\frac{d a_{D,n}(t)}{dt}&=&-i(\epsilon_D+n\omega_0)a_{D,n}(t)-iV_0a_{T,n}(t), \nonumber \\
\frac{d a_{T,n}(t)}{dt}&=&-i(\epsilon_T+n\omega_0)a_{T,n}(t)-iV_0a_{D,n}(t)-i \lambda \sqrt{n}a_{T,n-1}(t)-i \lambda \sqrt{n+1}a_{T,n+1}(t). 
\nonumber \\
\label{equ-t}
\end{eqnarray}

 Equations (\ref{equ-t}), that are exactly equivalent to Eqs. (\ref{equ-t-tilde}), are solved numerically for $t> 0$ with the initial condition that the trap is empty and no phonons are present: 
$a_{D,n}(t=0)=\delta_{n,0}$ and $a_{T,n}(t=0)=0 ,\; \forall n$. The time-dependent occupancy of the trap, that of the dot  and the number of phonons in the system are then calculated as

\begin{equation}
n_{T}(t)\equiv \langle \psi(t)|\hat{c}_{T}^{\dagger} \hat{c}_{T}|\psi(t) \rangle =\sum_{n=0}^{\infty}| a_{T,n}(t)|^2,
\label{nT2}
\end{equation}

\begin{equation}
n_{D}(t)\equiv \langle \psi(t)|\hat{c}_{D}^{\dagger} \hat{c}_{D}|\psi(t) \rangle =\sum_{n=0}^{\infty}| a_{D,n}(t)|^2,
\label{nD}
\end{equation}

and 
\begin{equation}
n_{b}(t)\equiv \langle \psi(t)|\hat{b}^{\dagger} \hat{b}|\psi(t) \rangle =\sum_{n=0}^{\infty} n(|a_{T,n}(t)|^2+|a_{D,n}(t)|^2).
\label{nb2}
\end{equation}
respectively. 

Having $n_T(t)$, we calculate the time-averaged trap occupancy  as

\begin{equation}
\langle n_{T} \rangle =\frac{1}{t_{max}} \int_{0}^{t_{max}} dt\; n_T(t),
\label{nT-taver}
\end{equation}
where $t_{max}$ is the maximum value of time in our calculations (usually $\omega_{0}t_{max}=1000$). It will be apparent from the results that we will present next that much smaller values of $t_{max}$ could be used to give an accurate value of $\langle n_{T} \rangle$. $\langle n_{T} \rangle$ is the physically relevant occupancy of the trap for these slower processes competing with hopping and electron-phonon interaction in the filling of the trap, such as light emission. 

\subsection{ A four-sites system}
\label{sec4sitestheory}

In this subsection the core of the QD is modeled by a system of three sites and, consequently, three energy bands. The sites, labeled 1, 2 and 3, have a single energy level $\epsilon_D$, connected by a first neighbors hopping parameter $\gamma$, while the trap has a single energy level $\epsilon_T$ coupled to phonons. The Hamiltonian of the uncoupled dot and trap subsystems thus reads,

\begin{equation}
\hat H_0=\hat H_D+\hat H_T,
\label{Ham-3}
\end{equation}
where $\hat H_T$ is given by Eq. (\ref{Ham-T}) and the dot Hamiltonian is 

\begin{equation}
\hat H_D=\sum_{i=1}^{3}\epsilon_D \hat n_i+ \gamma \hat c_{1}^{\dagger} \hat c_{2}+\gamma \hat c_{2}^{\dagger} \hat c_{3}+h.c. ,
\label{Ham4-D}
\end{equation}
where $\hat c_{i}$ ($\hat c_{i}^{\dagger}$) are the annihilation (creation) operators for electrons at the dot site $i$ and $\hat n_i$ is the number operator .

The dot's site 3 is coupled to the trap by a hopping parameter $V_0$, the hopping Hamiltonian being

\begin{equation} 
\hat H_{hop}= V_0(\hat c_{3}^{\dagger} \hat c_{T} +\hat c_{T}^{\dagger} \hat c_{3}). 
\label{H-int4}
\end{equation}

The time-dependent problem is solved in the same way as for the two-sites system.  We chose a site-basis of states that are written in an occupation number representation as 

\begin{eqnarray}
|\varphi_{1,n}\rangle &= &|1 0 0 0\rangle \otimes|n \rangle,  \nonumber \\
|\varphi_{2,n}\rangle &= &|0 1 0 0\rangle \otimes|n \rangle,  \nonumber \\
|\varphi_{3,n}\rangle &= &|0 0 1 0\rangle \otimes|n \rangle,  \nonumber \\
|\varphi_{T,n}\rangle &= &|0 0 0 1\rangle \otimes|n \rangle,  \nonumber \\ 
\label{basis-4}
\end{eqnarray}
with $n=0,1,2...$.  The time-dependent wave function is then written as a linear combination of the basis states

\begin{equation}
| \psi(t) \rangle= \sum_{n=0}^{\infty}[ a_{T,n}(t)|\varphi_{T,n}\rangle+ \sum_{i=1}^{3} a_{i,n}(t)|\varphi_{i,n}\rangle ],
\label{psi3-t}
\end{equation}
and the time-dependent Sch\"ordinger equation is projected onto the basis, leading to the following system of coupled linear differential equations

\begin{eqnarray}
\frac{d a_{1 ,n}(t)}{dt}&=&-i(\epsilon_D+n\omega_0)a_{1,n}(t)-i \gamma a_{2,n}(t),  \nonumber \\
\frac{d a_{2 ,n}(t)}{dt}&=&-i(\epsilon_D+n\omega_0)a_{2,n}(t)-i \gamma a_{1,n}(t)-i \gamma a_{3,n}(t),  \nonumber \\
\frac{d a_{3 ,n}(t)}{dt}&=&-i(\epsilon_D+n\omega_0)a_{3,n}(t)-i \gamma a_{2,n}(t)- iV_0 a_{T,n}(t), \nonumber \\
\frac{d a_{T,n}(t)}{dt}&=&-i(\epsilon_T+n\omega_0)a_{T,n}(t)-iV_0 a_{3,n}(t)-i \lambda \sqrt{n} a_{T,n-1}(t)-i \lambda \sqrt{n+1} a_{T,n+1}(t). 
\nonumber \\
\label{equ3-t}
\end{eqnarray}

Now we need to specify the initial conditions. 
The dot Hamiltonian of Eq. (\ref{Ham4-D}) is diagonalized producing three band states.
The ground state and the first and second excited states have energies $e_0=\epsilon_D-\gamma \sqrt{2}$, $e_1=\epsilon_D$ and $e_2=\epsilon_D+\gamma \sqrt{2}$ with eigenstates

\begin{eqnarray}
|\phi_0\rangle& =& \frac{1}{2}|100\rangle -\frac{1}{\sqrt{2}} |010\rangle+\frac{1}{2}|001\rangle,  \nonumber \\
|\phi_1\rangle &=& \frac{1}{\sqrt{2}}|100\rangle -\frac{1}{\sqrt{2}}|001\rangle, \nonumber \\
|\phi_2\rangle& =& \frac{1}{2}|100\rangle +\frac{1}{\sqrt{2}} |010\rangle+\frac{1}{2}|001\rangle, \nonumber \\
\label{phi-band}
\end{eqnarray}
respectively.  We assume that at $t=0$ the electron is in one of these band states with zero phonons in the system. 
Consequently  $a_{1,n=0}(t=0)$, $a_{2,n=0}(t=0)$ and $a_{3,n=0}(t=0)$ will equal the  weights of the states $|100\rangle$, $|010\rangle$ and
$|001\rangle$ in the corresponding wave function of Eq. (\ref{phi-band}), the rest of the coefficients will be zero and $a_{T,n}(t=0)=0 \; \forall n$. 

The occupancies of the trap and dot and the number of phonons are

\begin{equation}
n_{T}(t)=\sum_{n=0}^{\infty}| a_{T,n}(t)|^2,
\end{equation}

\begin{equation}
n_{D}(t) =\sum_{n=0}^{\infty} \sum_{i=1}^{3} | a_{i,n}(t)|^2,
\end{equation}

and 
\begin{equation}
n_{b}(t) =\sum_{n=0}^{\infty} n[|a_{T,n}(t)|^2+\sum_{i=1}^{3} |a_{i,n}(t)|^2],
\end{equation}
respectively.  These quantities will additionally be labeled by subindexes 0, 1 or 2 for calculations in which the initial electronic band state is $e_0$, $e_1$ or $e_2$, respectively. If we assume that the electron has the same probability of initially being in any of the three states, the magnitude of interest is the mean occupation of the trap 
$n_{Tav}(t)=\frac{1}{3}(n_{T0}(t)+n_{T1}(t)+n_{T2}(t))$. We also calculate the time-averaged values of the trap occupancies as in Eq.(\ref{nT-taver}).

\section{Results}
\label{sec:results}
\subsection{Two-sites system}

In our calculations we take $\omega_0$ as unit of energy, $\epsilon_D=0$,  and keep the value of $\tilde \epsilon_T$ fixed for any $\lambda$, in order to compare results for different values of $\lambda$. For ZnO quantum dots $\omega_0 \simeq 35$ meV ($\omega_0=53 $ ps$^{-1}$)\cite{Janotti_PRL, Janotti_APL} and typical values of $|\tilde \epsilon_T|$ are about 20 $\omega_0$ \cite{JPCB_104_1715_vanDijken,JPCC_116_20633_cohn}. This value is large enough so that we have to include a large number of phonons in the calculation. Therefore we take $\tilde \epsilon_T \simeq -10 \omega_0$.
Nevertheless we have checked that for $\tilde \epsilon_T \simeq -20 \omega_0$ the behavior of the trap occupancy as a function of time that we are going to present in this work remains the same but at larger values of $\lambda$. 
The time dependent Eqs. (\ref{equ-t}) are solved using a standard Runge-Kutta method and $n_T(t)$ and $n_D(t)$ are then calculated by Eqs. (23) and (24), respectively.  The value of the time step depends on the parameters and the calculations are checked by ensuring that the normalization condition $n_{T}(t)+n_{D}(t)=1$ is fulfilled with an accuracy better than 1$\%$,  $ \forall t$. Therefore only the trap occupancy $n_T(t)$ will be shown in this section. The number of phonons that should be included in the calculation obviously depends on $\lambda$. We find that even if $n_b(t)$ gets smaller than 1, one needs to include at least 40 phonons  for $\lambda \leq \omega_0$ and more that 100 phonons for $\lambda \gg \omega_0$. For $n_b(t)$ on the order of some tens, one needs to include typically 3-4$n_b$. This indicates that a large number of phonons are virtually involved in the process.
The calculations run up to a maximum time $\omega_0 t_{max}=1000$ which is usually enough to have converged results. 

As we will see below, our calculated $n_T(t)$ is an oscillatory function of time that may be perfectly periodic or not depending on the parameters $\lambda$, $V_0$ and $\tilde \epsilon_T$. In many cases we can identify two types of oscillations that will be named "fast" and "slow" oscillations.  Although they are not periodic sensu stricto, the distance between relative maxima or minima are approximately constant in time. The fast oscillations originate from the electronic hopping between the dot level and the trap sublevel where the density of states is maximun $n \simeq g$, with hopping parameter $V_0$. The assignment is made because the calculated period is in very good agreement with Rabi's formula 
$T_R(n=g)=\pi/ \sqrt{V_0^2+(\tilde \epsilon_T+g \omega_0-\epsilon_D)^2/4}$. The slow oscillations originate from the electronic hopping between the dot level and a resonant or quasi-resonant sublevel of the trap, with a renormalized hopping parameter $\tilde V(n_r)$ or $\tilde V(n_{qr})$, respectively, given by Eq.  (\ref{V-tilde}). The assignment is made because the dependence of the calculated period with $\lambda$ follows the trends of Eq. (\ref{T-tilde}). Fast and slow oscillations, that are characteristics of the weak coupling ($\lambda \ll \omega_0$) and strong coupling ($\lambda \gg \omega_0$) regimes, respectively, may appear alone or together depending on other parameters and set the dynamical time scales. However, in the region of parameters $\lambda \geq \omega_0$ and $\lambda \simeq V_0$, where electron-phonon interaction and hopping compete, the features in $n_T(t)$ cannot be so easily assigned. We will also present results for the total number of excited phonons, Eq. (\ref{nb2}). It is important to point out that, in all the cases investigated in this work, this number is practically given by the first term on the right hand side of Eq. (\ref{nb2}). Although states with an electron and $n$ phonons at the dot are included in the basis set of Eq. (\ref{basis}) and in the calculation, they occur with very low probability in our final solution. Hence phonons are localized at the trap site and do not accompany the electron hop to the dot.
Since $n_T(t)$ and $n_b(t)$ are quasi-periodic functions of time, for clarity we will present only intervals of the full calculation in some cases. We can distinguish different regimes of the parameters.

a) In the weak coupling regime $\lambda \leq \omega_0$, the trap density of states is strongly peaked at $n \approx g \approx 0,1$ and we find basically fast oscillations with a period smaller than typically $\frac{1}{\omega_0}$.
However, for $\lambda \simeq \omega_0$ there are "bunches" of oscillations of large amplitude around the mean value of $n_T$ separated in time by the phonon period $T_0=\frac{2\pi}{\omega_0}$.
This is shown in Figure 1 where we plot $n_T(t)$ and $n_b(t)$ for $\tilde \epsilon_T= -10 \omega_0$, $\lambda=\omega_0$ and $V_0=0.5 \omega_0$. Since $T_R \ll T_0$, the phonons do not follow the quick electron hopping and $n_b(t)$ presents a perfect oscillatory behavior of period $T_0$. 
The large amplitude "bunches"  appear at times when the number of phonons is the smallest while the small amplitude oscillations appear at times when the number of phonons is the highest and  the electron mainly "stays" in the trap. This kind of behavior is seen for all values of $\tilde \epsilon_T$ and for all values of $V_0$, smaller or larger than $\omega_0$ although it shows up at higher values of $\lambda$ for increasing $V_0$. The perfect periodicity shown in Fig. 1 extends up to $t_{max}$. Note that the time averaged values of $n_T$ can be obtained at times $\omega_0 t\simeq 5-10$ (or $t \simeq 0.1-0.2$ ps
for $\omega_0=50$ ps$^{-1}$).
\begin{figure}[htbp]
\centering
\includegraphics[width=8cm]{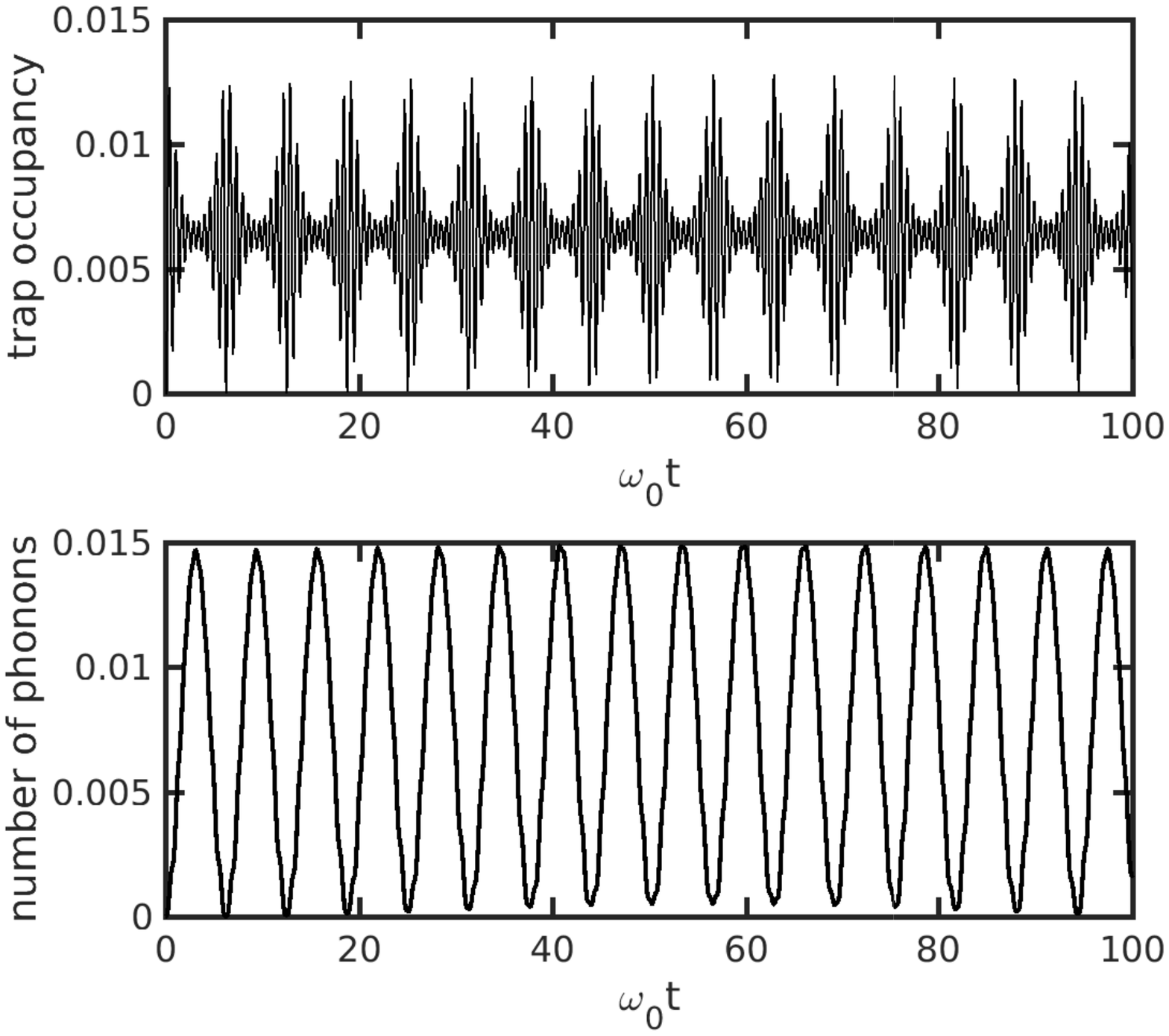}
\caption{The trap occupancy $n_T(t)$ (upper panel) and the number of phonons $n_b(t)$ (lower panel)as a function of time for 
$\tilde \epsilon_T=-10 \omega_0$, $V_0=0.5 \omega_0$ and $\lambda= \omega_0$.} 
\label{fig1}
\end{figure}

b) Intermediate coupling regime. For $\lambda > \omega_0$ the weight of the trap density of states moves to sublevels of higher energy and also becomes wider. Hence $n_T(t)$ depends strongly on $V_0$.

b1) If $V_0 \leq \omega_0$, basically only the trap sublevel nearest in energy to $\epsilon_D$ is available for electrons to hop to. 
If there is one trap sublevel resonant with the dot level, the electron transfer proceeds via this sublevel at relatively small values of $\lambda$, with the consequent increase in the values of $n_T(t)$. 
This is illustrated in Figure 2 where the parameters are the same as in Fig. 1 except that we have increased $\lambda$ to 
$\lambda=1.5 \omega_0$. The bunches of hops to the $n \approx g=2.25$ sublevel are still visible and are superimposed to another oscillation of much larger period. This is the slow oscillation associated to the resonant electron transfer to the sublevel with $n_r=10$ given by Eq. (\ref{T-tilde}). 
The full electron dynamics is set after $\omega_0 t\simeq 100$ (or $t \simeq 2$ ps for $\omega_0=50$ ps$^{-1}$).
It is also worth noticing in this Figure that the the trap occupancy is above 1$\%$ at almost any time. With increasing $\lambda$ the amplitude of the fast oscillations gets smaller and eventually disappears (see Fig. 4, black line), while the amplitude of the slow oscillations increases as the strong coupling behavior of Eq. (\ref{nT-res}) is approached. 
If there is no resonant sublevel, the electron transfer proceeds via the quasi-resonant sublevel nearest in energy to $\epsilon_D$ and we find the same kind of behavior as for the resonant sublevel: $n_T(t)$ oscillates with a period given approximately by Eq. (\ref{T-tilde}) and its amplitude is smaller than for the resonant case. Perfect periodicity up to $t_{max}$ is also found.
\begin{figure}
\centering
\includegraphics[width=8cm]{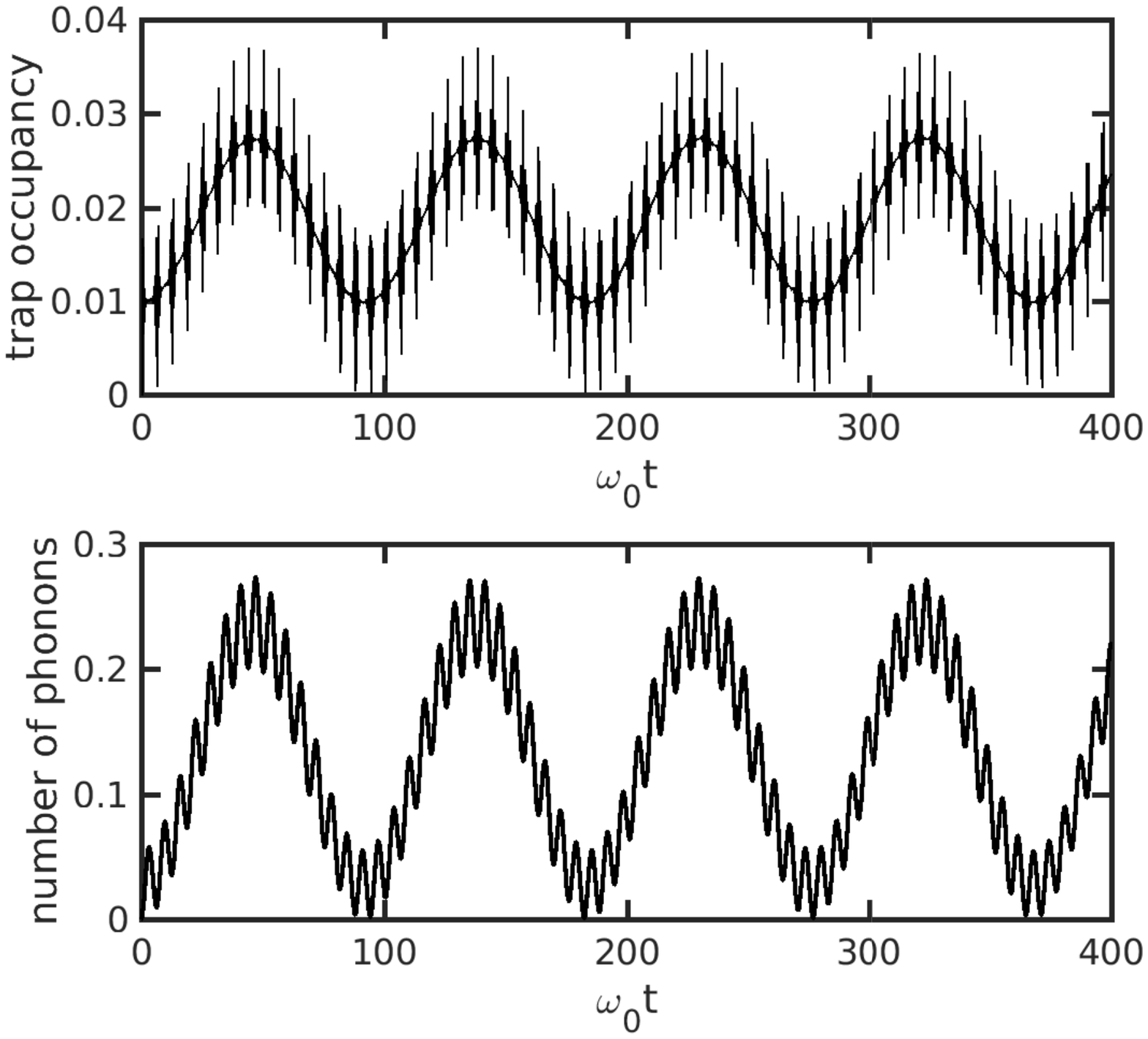}
\caption{The trap occupancy $n_T(t)$ (upper panel) and the number of phonons $n_b(t)$ (lower panel) as a function of time for 
$\tilde \epsilon_T=-10 \omega_0$, $V_0=0.5 \omega_0$ and $\lambda= 1.5 \omega_0$.}
\label{fig2}
\end{figure} 

An interesting case is the one with $\tilde \epsilon_T=-10.5 \omega_0$ in which two trap sublevels are above and below $\epsilon_D$ at the same energetic distance, illustrated in Figure 3 for $V_0=0.5 \omega_0$ and $\lambda=2.5 \omega_0$ ($g=6.25$). $n_T(t)$ shows the fast oscillations, that do not show up in $n_b(t)$, superimposed to other non-periodic oscillations of larger amplitude. These non-periodic oscillations come from the superposition of electron hops to sublevels with $n=10$ and $n=11$ that for this $\lambda$ have rather different amplitudes and frequencies. 
\begin{figure}
\centering
\includegraphics[width=8cm]{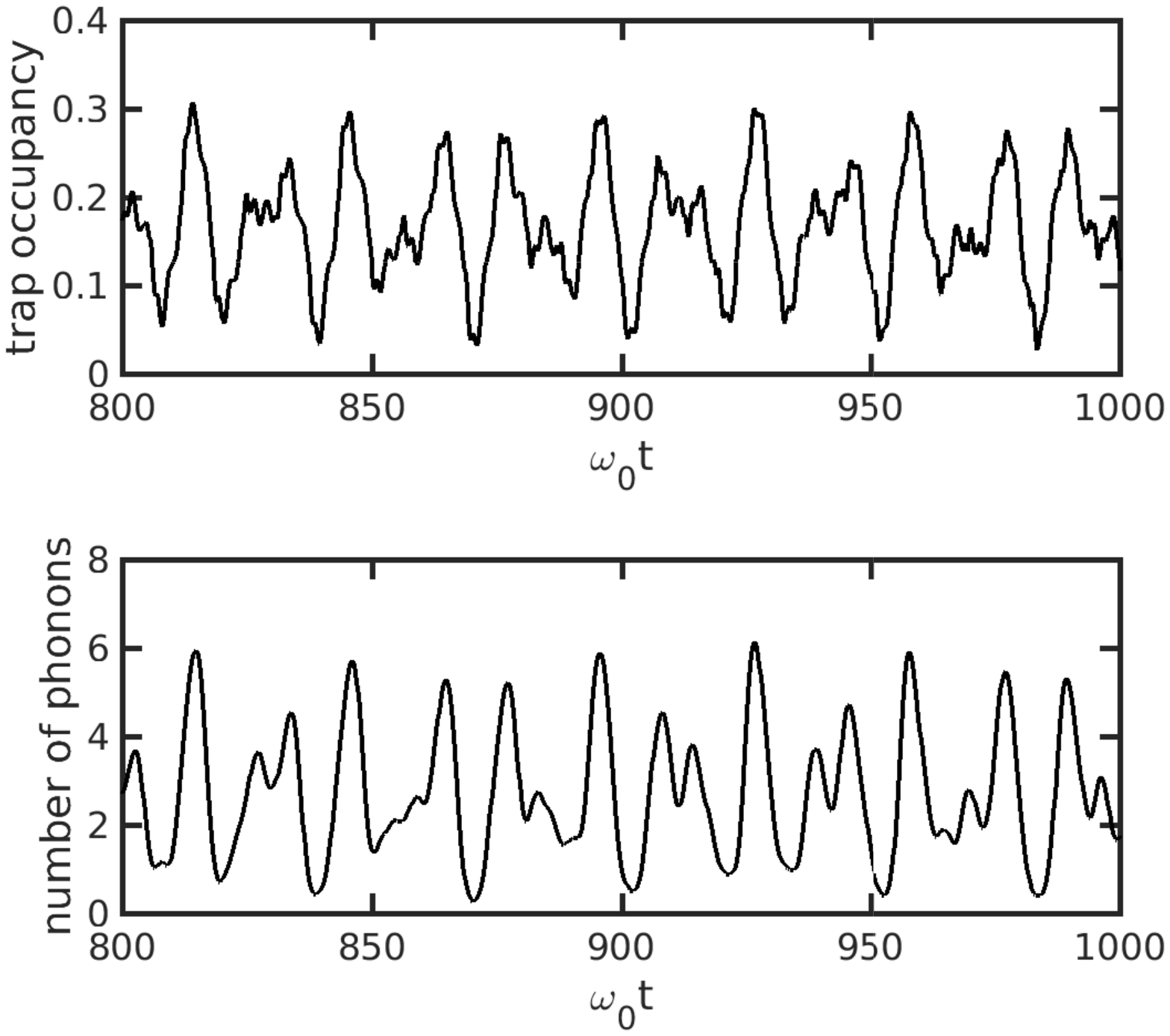}
\caption{The trap occupancy $n_T(t)$ (upper panel) and the number of phonons $n_b(t)$ (lower panel) as a function of time for 
$\tilde \epsilon_T=-10.5 \omega_0$, $V_0=0.5 \omega_0$ and $\lambda= 2.5 \omega_0$.}
\label{fig3}
\end{figure}

b2) For $\lambda > \omega_0$ and $V_0 \geq \omega_0$ several trap sublevels enter into play. In general, we find the fast oscillations superimposed to other features coming from other sublevels. The relative contributions of both kind of features depend on $\lambda$ and $V_0$

 Figure 4 compares the cases with $V_0= 0.5 \omega_0$ (black), $V_0= \omega_0$ (red) and $V_0=2 \omega_0$ (blue), 
for $\tilde \epsilon_T = -10 \omega_0$ and  $\lambda=2 \omega_0$.
For $V_0= 0.5 \omega_0$ and $V_0= \omega_0$ we find the same oscillatory behavior of Fig. 2, with fast oscillations corresponding to $n=4$ superimposed to slow ones corresponding to the resonant trap sublevel $n_r=10$. There are important differences, however. The case with $V_0= \omega_0$ shows the formation of plateaux of width $ \omega_0 t \simeq 10$ at the maxima and/or minima of $n_T(t)$, around which the fast electronic hops take place,
and the perfect periodicity is lost in $n_T(t)$ as well as in $n_b(t)$. 
This is due to the fact that for $V_0= 0.5 \omega_0$  only the resonant sublevel $n_r=10$ is basically accessible while at least two more sublevels enter in competition with the resonant one for $V_0= \omega_0$. Also, contrary to intuition, a larger value of the hopping parameter decreases the average occupancy of the trap. Increasing this value to $V_0=2 \omega_0=\lambda$ enlarges the amplitude of the fast oscillations and dilutes the effect of the resonant sublevel to such an extent that it cannot be clearly appreciated. At variance from the other cases, the number of phonons does not follow the trap occupancy as this number has relative maxima/minima separated by the phonon period $T_0$. Therefore, electrons and phonons are completely separated subsystems in this case.
\begin{figure}
\centering
\includegraphics[width=8cm]{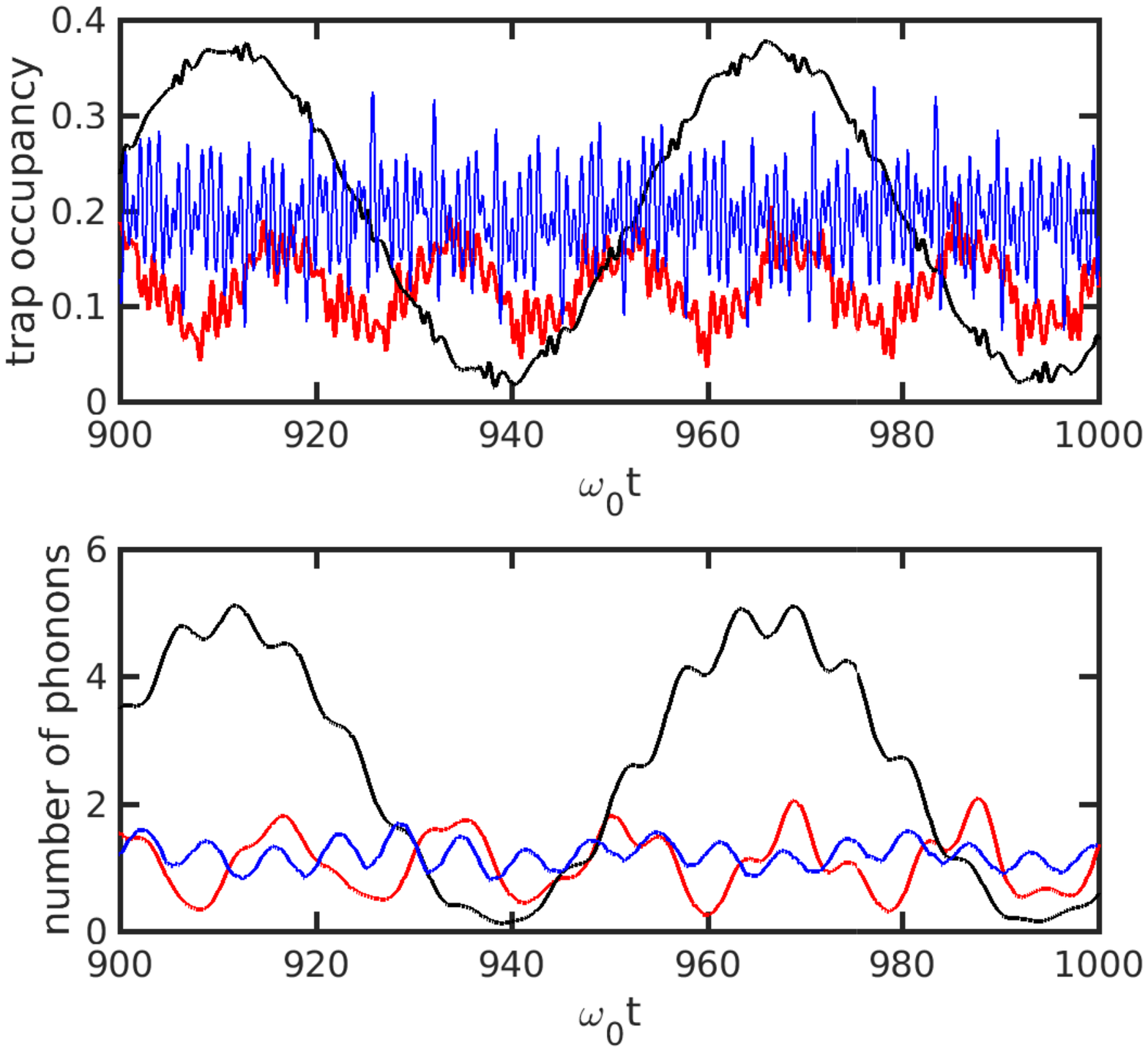}
\caption{The trap occupancy $n_T(t)$ (upper panel) and the number of phonons $n_b(t)$ (lower panel) as a function of time for 
$\tilde \epsilon_T=-10 \omega_0$, $\lambda= 2 \omega_0$ and $V_0=0.5 \omega_0$ (black lines), $V_0= \omega_0$ (red lines) 
and $V_0=2 \omega_0$ (blue lines).}
\label{fig4}
\end{figure}

In Figure 5 we increase the electron-phonon interaction to $\lambda=3.3 \omega_0$ while keeping $V_0=2 \omega_0$ and $\tilde \epsilon_T = -10 \omega_0$. 
For clarity, we show the results in the time span $400 \leq \omega_0 t \leq 500$ but this pattern repeats at other intervals along the calculation, as well.
We find intervals where there are kind of plateaux of width $ \omega_0 t \simeq 5$ with fast oscillations followed by rapid drops in the occupancy (as for $ \omega_0 t \ge 450 $ in this Figure)
separated by other intervals of width $ \omega_0 t \simeq 20-30$ where the oscillations in $n_T$ around its mean value are considerably smaller
 (as for $400 \leq \omega_0 t \leq 420$). This behavior is reminiscent of blinking although with a very different time scale.
Notice here that the trap is at least 20$\%$ occupied at any time. 
The number of phonons is a more regular function showing the same behavior and its period is probably related to the resonant transition.
This kind of behavior is obtained when the hopping and electron-phonon interaction compete, that is, for $\lambda \geq V_0$. 
\begin{figure}
\centering
\includegraphics[width=8cm]{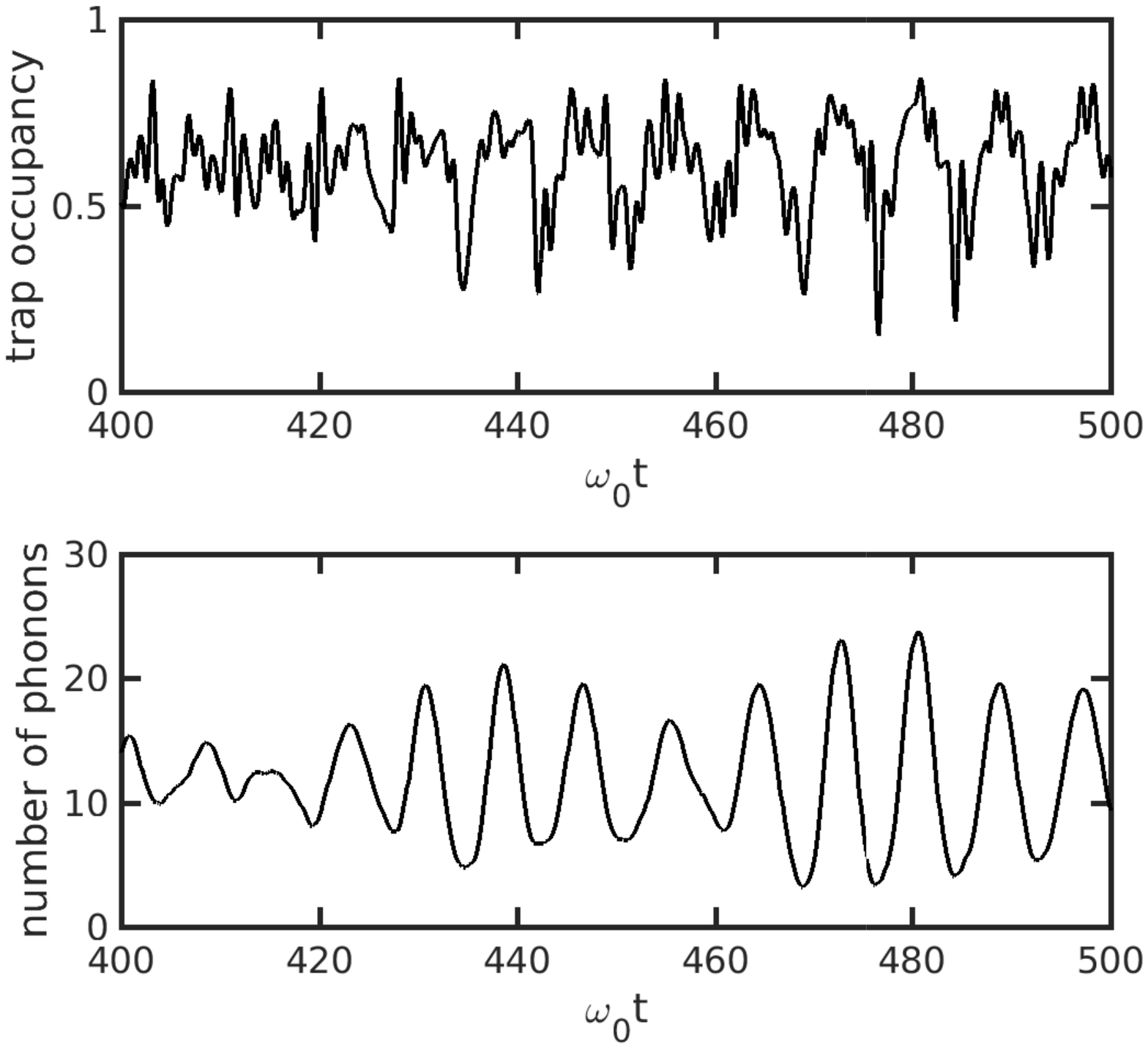}
\caption{The trap occupancy $n_T(t)$ (upper panel) and the number of phonons $n_b(t)$ (lower panel) as a function of time for 
$\tilde \epsilon_T=-10 \omega_0$, $V_0=2 \omega_0$ and $\lambda= 3.3 \omega_0$.} 
\label{fig5}
\end{figure}

c) Strong coupling regime $\lambda \gg V_0$ and $\lambda \gg \omega_0$. This is the regime dictated by Eqs. (\ref{nT-res}) or (\ref{nT-nores}) and we recover the perfect periodic oscillatory behavior predicted by these equations. Here we show the case with $\tilde \epsilon_T=-10.5 \omega_0$, $V_0=0.5 \omega_0$ 
and $ \lambda=3.3 \omega_0$ ($g=10.89$), for which the renormalized hopping parameters to the trap sublevels with $n=10$ and $n=11$ are very similar and we find the coherent superposition of two oscillatory motions of almost the same amplitude and similar frequencies. Actually, the oscillations of this Figure have a period $\simeq \tilde T(n=10) \simeq \tilde T(n=11)$. 
\begin{figure}
\centering
\includegraphics[width=8cm]{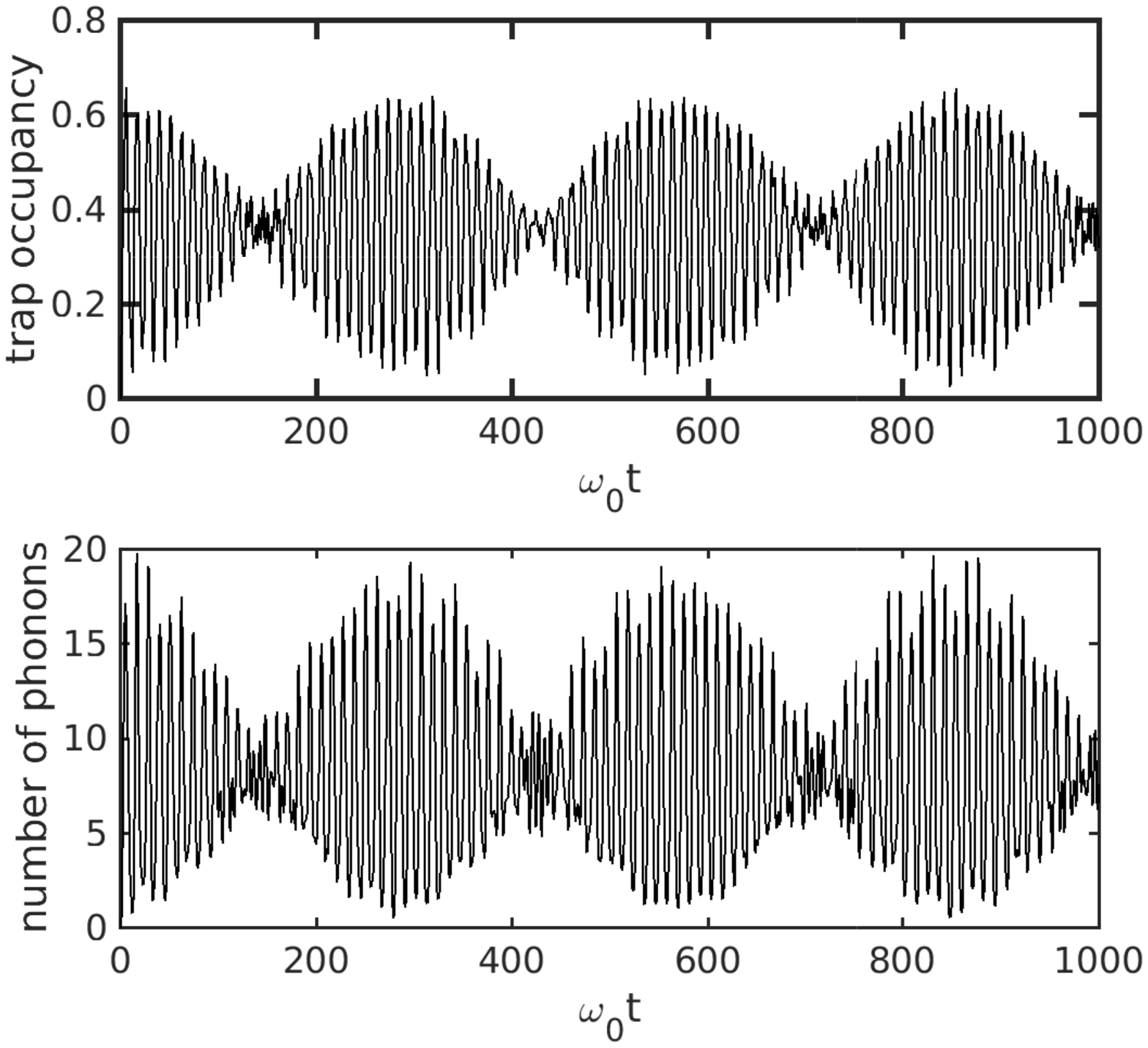}
\caption{The trap occupancy $n_T(t)$ (upper panel) and the number of phonons $n_b(t)$ (lower panel) as a function of time for 
$\tilde \epsilon_T=-10.5 \omega_0$, $V_0=0.5 \omega_0$ and $\lambda= 3.3 \omega_0$.}
\label{fig6}
\end{figure}

In almost all the cases investigated in this work, changing $\tilde \epsilon_T$, $\lambda$ and $V_0$, the system dynamics is established in a very short time, as seen in Figs. 1, 2 and 6. The exception are the cases with $V_0\simeq \lambda \geq \omega_0$ where we find a long time transient. 
An example is $\tilde \epsilon_T = -10 \omega_0$ and  $\lambda=V_0=2 \omega_0$, the long-term dynamics of which is shown by the blue line of Fig. 4. Up to $\omega_0 t \simeq 100$ the dynamics is more similar to the 
weak coupling regime of Fig. 1, with bunches of large amplitude fast oscillations around the mean value separated by $T_0$, but as time goes on the separation gets shorter and shorter so that for $\omega_0 t \geq 500$ we already obtain the behavior shown in Fig. 4. We interpret this long transient dynamics as the effect of a strong competition between hopping and electron-phonon interactions. We should stress that the average trap occupancy is set on the short-term and does not change with time, the only change is in the separation of the bunches of oscillations around the average value.

Perhaps the most striking effect of electron-phonon interaction in this problem is the formation of resonances. They show up in $n_T(t)$ and $n_b(t)$ as a perfect periodic motion of very large amplitude and period that, for given $\tilde \epsilon_T$ and $V_0$, only occurs at particular values of $\lambda$: a change in $\lambda$ of less that 10$\%$ kills the resonance. Figure 7 depicts this motion for  $V_0=1.6 \omega_0$, $\lambda=2.2 \omega_0$ and 
$\tilde \epsilon_T=-13.4 \omega_0$, a value of $\tilde \epsilon_T$ for which there is no trap sublevel in resonance with $\epsilon_D$ and slow oscillations should be more similar to the ones of Fig. 3. This is the picture of an electron that is trapped and detrapped each $\simeq \frac{100}{\omega_0}$ s (or 2 ps for $\omega_0=50$ ps$^{-1}$) but this time can be shorter or even 5 times larger in other cases.
We find resonances for all the values of $\tilde \epsilon_T$ investigated in this work and for all values of $V_0$ as long as $V_0 >\omega_0$, which indicates that several trap sublevels coherently cooperate in their formation. 
\begin{figure}
\centering
\includegraphics[width=8cm]{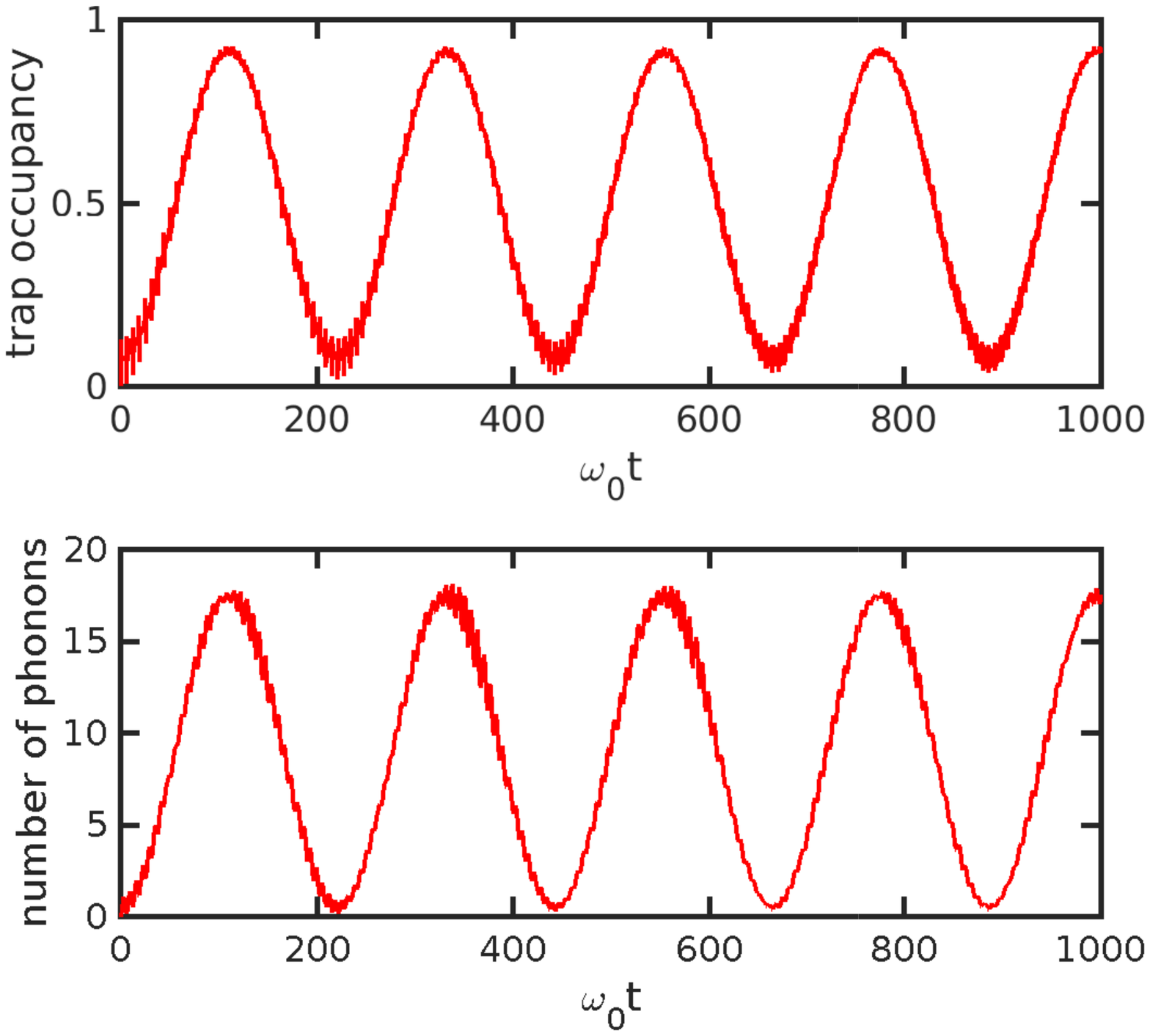}
\caption{The trap occupancy $n_T(t)$ (upper panel) and the number of phonons $n_b(t)$ (lower panel) as a function of time for 
$\tilde \epsilon_T=-13.4 \omega_0$, $V_0=1.6 \omega_0$ and $\lambda= 2.2 \omega_0$.}
\label{fig7}
\end{figure}

The resonances manifest as spikes in the time-averaged 
trap occupancy $\langle n_T \rangle (\lambda)$ with values larger than 40 $\%$. Figure 8 shows 
$\langle n_T \rangle (\lambda)$ for $\tilde \epsilon_T=-10 \omega_0$ (left panel) and $\tilde \epsilon_T=-13.4 \omega_0$ (right panel) for several values of $V_0 \ge \omega_0$. For $V_0=\omega_0$ (black symbols)
$\langle n_T \rangle (\lambda)$ is a smooth function having a broad maximum near $\lambda \simeq 3.3$ for which the maximum of the trap density of states is quasi-resonant with $\epsilon_D$. $\langle n_T \rangle (\lambda)$ is larger than 0.5 because $n_T(t)> 0.2$ $\forall t$ (see Fig. 5).
Concentrating first in the case $\tilde \epsilon_T=-10 \omega_0$, we note that for $V_0=1.2 \omega_0$ (orange dots) the broad maximum of $\langle n_T \rangle (\lambda)$ is split into a maximum  and a shoulder occurring at higher and smaller values of $\lambda$ respectively. By increasing $V_0$ to $V_0=1.4 \omega_0$ (magenta dots) the shoulder has evolved into a peak that moves to smaller values of $\lambda$ and
narrows as $V_0$ increases. Finally, for $V_0= 2 \omega_0$ (green dots) the spike is at $\lambda= 1.5 \omega_0$. 
To show that resonances are formed independently of $\tilde \epsilon_T$ the right panel of Fig. 8 shows the results for $\tilde \epsilon_T=-13.4 \omega_0$.
For $V_0= \omega_0$ (black squares)
$\langle n_T \rangle (\lambda)$ is a smooth function having a broad maximum near $\lambda \simeq 3.5$ and, with increasing $V_0$, the maximum is displaced toward higher values of $\lambda$ while two peaks develop
 that are clearly visible for $V_0= 1.6 \omega_0$ and $V_0= 1.8 \omega_0$, and  even three peaks appear for $V_0= 2\omega_0$.
We have never found spikes below $\lambda= 1.5 \omega_0$, indicating that the electron-phonon interaction has to be strong enough to form these resonances.
Additionally, for $V_0= 2 \omega_0$ and larger other small peaks develop at $\lambda \simeq 3.2 \omega_0$ that also follows the trend of moving 
to smaller $\lambda$ for increasing $V_0$.
It is worth noticing in Fig. 8 that the maximum value of the occupancy is $\geq$ 60$\%$ independently of $V_0$ (as long as $V_0 \geq \omega_0$) and of $\tilde \epsilon_T$. However $\langle n_T \rangle (\frac{\lambda}{\omega_0}\gg 1) \rightarrow 0.5$ for $\tilde \epsilon_T= -10\omega_0$ while 
$\langle n_T \rangle (\frac{\lambda}{\omega_0}\gg 1) \rightarrow 0$ for $\tilde \epsilon_T= -13.4\omega_0$, as predicted by Eqs. (\ref{nT-res}) and 
(\ref{nT-nores}) for cases where there exists or not a resonant trap sublevel, respectively. We will see in the next subsection that this different limiting 
behavior disappears when the dot is described by more than a single electronic level.
\begin{figure}
\centering
\includegraphics[width=8cm]{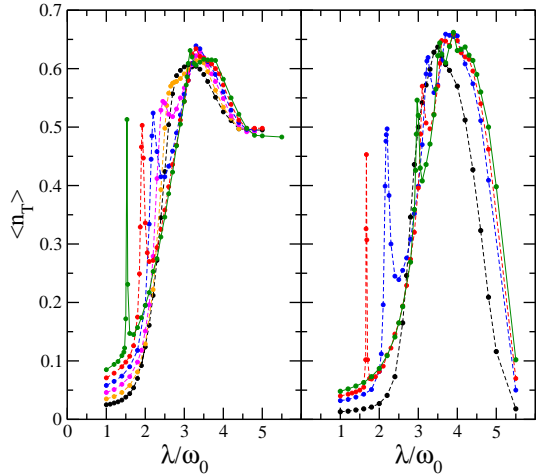}
\caption{The time-averaged trap occupancy $\langle n_T \rangle$ as a function of $\lambda$. Left panel is for
$\tilde \epsilon_T=-10 \omega_0$, and $V_0=\omega_0$ (black dots), $V_0=1.2 \omega_0$ (orange dots), $V_0=1.4 \omega_0$ (magenta dots),
$V_0=1.6 \omega_0$ (blue dots), $V_0=1.8 \omega_0$ (red dots) and $V_0=2 \omega_0$ (green dots).
Right panel is for $\tilde \epsilon_T=-13.4 \omega_0$, and $V_0=\omega_0$ (black dots),
$V_0=1.6 \omega_0$ (blue dots), $V_0=1.8 \omega_0$ (red dots) and $V_0=2 \omega_0$ (green dots).}
\label{fig8}
\end{figure}

\subsection{Four-sites system} 
\label{sec4sitesresults}

The ground state of the dot is now our zero of energy, $e_0=0$, so that $\epsilon_D=\gamma \sqrt{2}$ and the distance between electronic levels is $\Delta e=\gamma \sqrt{2}$.
The values of $\gamma$ will be chosen as to give values of $\Delta e$ similar to those of nanospheres of different sizes.
Considering nanospheres of radius $R$ as a simple model for quantum dots, the energy levels depend on $R$ as
$e_{ln}=\frac{\rho_{ln}^2}{2 m_{e}^{\ast} R^2}$, where $\rho_{ln}$ 
are the zeros of the spherical Bessel function of order $l$, $j_{l}(\rho_{ln})=0$, and $m_{e}^{\ast}$ is the effective electron mass. 
Typical values for ZnO are $m_{e}^{\ast}= 0.28 m_e$ and $\omega_0= 35$ meV, hence the average distances between the first three levels 
of nanospheres of radii between 2 and 8 nm, range in between 8$\omega_0$ and 0.5$\omega_0$, respectively. We will take $\gamma$ as to give values of  
$\Delta e$ in the same range. 
The rest of the parameters are as for the two-sites system and the calculations are preformed in the same way. The time dependent Eqs. (32) are solved using a standard Runge-Kutta method and $n_T(t)$ and $n_D(t)$ are then calculated by Eqs. (34) and (35), respectively, ensuring that the normalization condition $n_{T}(t)+n_{D}(t)=1$ is fulfilled with an accuracy better than 1$\%$,  $ \forall t$. Therefore only the trap occupancy $n_T(t)$ will be shown in this section. 

The results for the time-dependent trap occupancies $n_{T0}(t)$, $n_{T1}(t)$ and $n_{T2}(t)$ can be understood from the ones found for the two-sites system.
Although the time dynamics is obviously more involved, one can still distinguish the fast and slow oscillations associated to electron hopping between the trap and the last site of the chain where it is connected, that were described in the previous subsection. 
 In the weak coupling regime $\lambda \leq \omega_0$, we find fast oscillations that form bunches at $\lambda \simeq \omega_0$, independently of the rest of the parameters. When $V_0 \ll \Delta e$ each level $e_0$, $e_1$ or $e_2$ only couples to the trap so they are basically independent. The most interesting case is  $V_0 \geq \Delta e$ and $V_0\geq \omega_0$ for which two of the dot's levels can be coupled trough the trap and coherently involved in its filling.  In our opinion, the newest result to be presented in this subsection is the mean occupancy $n_{Tav}(t)$ because it can be rather time-independent when the oscillations of different $n_{Ti}(t)$ are out of phase. Two of these cases are presented in Fig.9. In the upper panel, $\tilde \epsilon_T=-10.5 \omega_0$, $\gamma=1.2 \omega_0$ ($\Delta e=1.7 \omega_0$), $V_0=2\omega_0$ and $\lambda=3.7 \omega_0$,
and we note that the oscillation of $n_{Tav}(t)$ around its time-averaged value is about 10$\%$ as maximum. Also note the rather flat parts of the curve
 $n_{T2}(t)$ that we also found for the two-sites system. In the lower panel of Fig. 9,  $\tilde \epsilon_T=-10\omega_0$, $\gamma=0.3 \omega_0$
($\Delta e=0.42 \omega_0$), $V_0=\omega_0$ and $\lambda=3.1 \omega_0$. In this case the oscillations of $n_{Tav}(t)$ around its time-averaged value are also small and 
it is remarkable that all the three $n_{T0}(t)$, $n_{T1}(t)$ and $n_{T2}(t)$ get similar values irrespective of the different relative position of 
$\tilde \epsilon_T$ with respect to $e_0$, $e_1$ and $e_2$, respectively. 
\begin{figure}
\centering
\includegraphics[width=8cm]{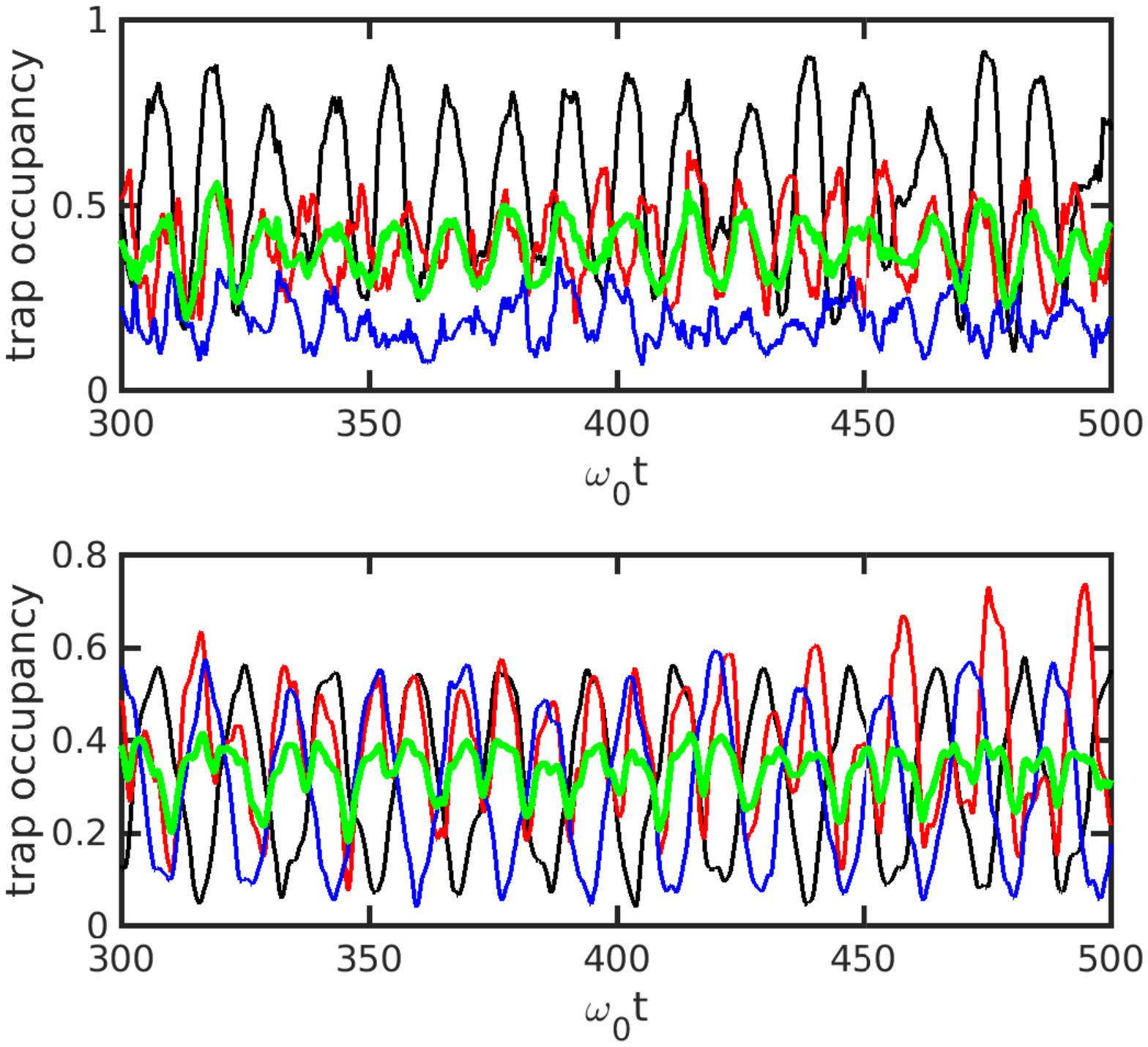}
\caption{The trap occupancies $ n_{T0}$ (black lines), $ n_{T1}$ (red lines), $ n_{T2}$ (blue lines)and $ n_{T,av}$ (green lines)
as a function of time. Upper panel: $\tilde \epsilon_T=-10.5 \omega_0$, $\gamma=1.2 \omega_0$, $V_0=2\omega_0$ and $\lambda=3.7 \omega_0$.
Lower panel: $\tilde \epsilon_T=-10\omega_0$, $\gamma=0.3 \omega_0$, $V_0=\omega_0$ and $\lambda=3.1 \omega_0$.  }
\label{fig9}
\end{figure}

Time-averaged trap occupancies $\langle n_{T0} \rangle$, $\langle n_{T1} \rangle$ and $\langle n_{T2} \rangle$ as a function of $\lambda$ for $\tilde \epsilon_T=-10 \omega_0$ and $\tilde \epsilon_T=-10.5 \omega_0$ and several values of $\gamma$ are presented in Figs. 10 and 11 for
$V_0=\omega_0$ and $V_0=2 \omega_0$, respectively. 
Our results in Fig. 10, where  $V_0=\omega_0$, point out to the mixing between two or more of the dot's levels trough the trap, that of course is more important for the smaller values of $\gamma$ but also happens up to our highest value. 
We first note the spikes and peak features described in the previous subsection that did not appear at $V_0=\omega_0$ for a two-sites system 
but that show up in most of the cases now, even when $\Delta e=3.4 V_0$. For comparison, the red open symbols in the bottom panels represent the results for a two-sites system having energy differences $e_D-\tilde \epsilon_T$ equal to $e_1-\tilde \epsilon_T$. Therefore, these resonances have to originate from the the coherent coupling of two dot's levels trough the trap. Moreover, the trap occupancies take similar values independently of whether the dot's level is resonant with a trap sublevel or not and, for the smaller values of $\gamma$, are also rather independent of the dot's level from which the electron started to hop, these facts being also indicative of the level coupling. 
We have to stress that this is not the behavior if $V_0 < \omega_0$ because in this case the occupancies 
$\langle n_{Ti} \rangle$ are very dependent on the energetic distance between $e_i$ and its nearest in energy trap sublevel. 
As an example for $V_0=0.5 \omega_0$ 
and $\tilde \epsilon_T=-10 \omega_0$, where the trap sublevel $n_r=10$ is resonant with $e_0$, we find a maximum value of $\langle n_{T0} \rangle$ 
of 0.42, while 
the maximum value of $\langle n_{T0} \rangle$ is 0.11 for $\tilde \epsilon_T=-10.5 \omega_0$, $\gamma=0.3 \omega_0$ in both cases.
This large difference is not seen in the top panel of Fig. 10. 
Another example is $V_0=0.5 \omega_0$, $\gamma=0.6 \omega_0$
and $\tilde \epsilon_T=-10.5 \omega_0$, where the maximum values of $\langle n_{T0} \rangle$, $\langle n_{T1} \rangle$ and $\langle n_{T2} \rangle$ 
are 0.08, 0.17 and 0.16 respectively which is completely different from the results for $V_0=\omega_0$ shown in the Figure, where the three values are very similar.
Finally, we note that in all the calculations of $\langle n_{T0} \rangle$ for $\tilde \epsilon_T=-10 \omega_0$ represented in the left panels, in which $e_0$ is in resonance with the $n_r=10$ sublevel of the trap, we find 
$\langle n_{T0} \rangle (\frac{\lambda}{\omega_0}\gg 1) \rightarrow 0$, contrary to what was found for the two-sites system. We conclude that this is also an effect of the coupling among different electronic levels of the dot.
\begin{figure}
\centering
\includegraphics[width=8cm]{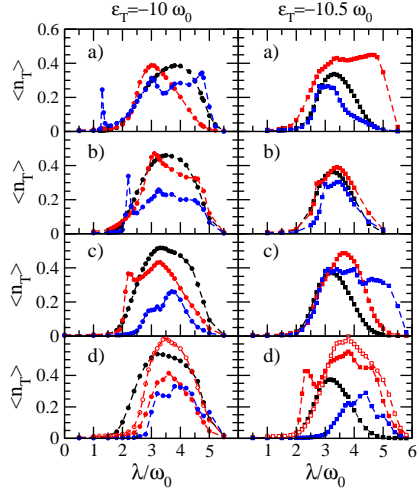}
\caption{Time-averaged trap occupancies $\langle n_{T0} \rangle$ (full black symbols), $\langle n_{T1} \rangle$ (full red symbols) and
$\langle n_{T2} \rangle$ 
(full blue symbols) as a function of $\lambda$ for $V_0=\omega_0$.
$\tilde \epsilon_T=-10 \omega_0$ (left panels) and 
$\tilde \epsilon_T=-10.5 \omega_0$ (right panels)
and the following values of $\gamma$ from top to bottom: a) $\gamma=0.3 \omega_0$ ($\Delta e=0.425 \omega_0$), b) 
$\gamma=0.6 \omega_0$ ($\Delta e=0.83 \omega_0$), c) $\gamma=1.2 \omega_0$ ($\Delta e=1.7 \omega_0$) and d) $\gamma=2.4 \omega_0$ ($\Delta e=3.4 \omega_0$).
The open symbols in the lower panels are the results for a two-sites system having the same energy differences.}
\label{fig10}
\end{figure}

 Figure 11 is as Fig. 10 but for $V_0=2 \omega_0$. The spikes and peak features are more pronounced than in Fig. 11 and, for the smaller values of $\gamma$, appear at the same value of $\lambda$ for $\langle n_{T0} \rangle$ , $\langle n_{T1} \rangle$ and $\langle n_{T2} \rangle$ this fact being also indicative of level mixing. It is remarkable that spikes show up not only at low but at high values of $\frac{\lambda}{\omega_0}$ as well. These spikes move toward higher values of $\lambda$, while the low $\frac{\lambda}{\omega_0}$ ones move toward lower 
values of $\lambda$ with increasing $V_0$. 
The comparison of the calculations of two-sites and four-sites systems presented in the lower panels shows additional structures in the later case.
\begin{figure}
\centering
\includegraphics[width=8cm]{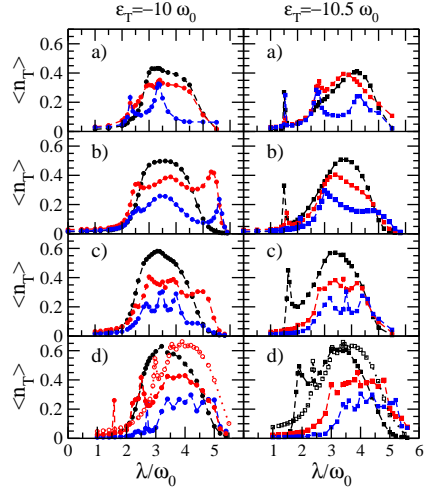}
\caption{Time-averaged trap occupancies $\langle n_{T0} \rangle$ (full black symbols), $\langle n_{T1} \rangle$ (full red symbols)
 and $\langle n_{T2} \rangle$ 
(full blue symbols) as a function of $\lambda$ for $V_0=2 \omega_0$.
$\tilde \epsilon_T=-10 \omega_0$ (left panels) and 
$\tilde \epsilon_T=-10.5 \omega_0$ (right panels)
and the following values of $\gamma$ from top to bottom: a)  $\gamma=0.3 \omega_0$ ($\Delta e=0.425 \omega_0$), b)
$\gamma=0.6 \omega_0$ ($\Delta e=0.83 \omega_0$), c) $\gamma=1.2 \omega_0$ ($\Delta e=1.7 \omega_0$) and d) $\gamma=2.4 \omega_0$ ($\Delta e=3.4 \omega_0$).
The open symbols in the lower panels are the results of a two-sites system having the same energy differences.}
\label{fig11}
\end{figure}

Our results are summarized in Figure 12 where we present the time-averaged mean trap occupancies $\langle n_{Tav} \rangle$ as a function of the electron-phonon coupling parameter $\lambda$ for $\tilde \epsilon_T=-10 \omega_0$ (left panels) and $\tilde \epsilon_T=-10.5 \omega_0$ (right panels), $V_0=\omega_0$ (top panels) and $V_0=2 \omega_0$ (bottom panels) and all the values of $\gamma$ investigated. The spikes and peaks structures found in practically all of the 
$\langle n_{Ti} \rangle$ shown in Figs. 10 and 11 are translated to $\langle n_{Tav} \rangle$ and these features are more prominent for $V_0=2 \omega_0$.
We note that, although each of the $\langle n_{Ti} \rangle$ differ to some extent depending on $\tilde \epsilon_T$, $\gamma$ and $V_0$, the mean occupancy 
$\langle n_{Tav} \rangle$ is rather independent of these parameters as long as $V_0\geq \omega_0$. Also note, that the maximum values of 
$\langle n_{Tav} \rangle$ are 30-40$\%$ in all the cases and the maximum is at $\frac{\lambda}{\omega_0} \simeq 3.5$ ($g \simeq 12$) for which the maximum of
the trap density of states is near $e_0$. This is not the case if $V_0<\omega_0$. As an  example for $V_0=0.5 \omega_0$ and
 $\gamma=0.6 \omega_0$ we obtain maximum values of $\langle n_{Tav} \rangle$ of 0.30 and 0.13 for $\tilde \epsilon_T=-10 \omega_0$ and 
 $\tilde \epsilon_T=-10.5 \omega_0$, respectively. Values of $g \simeq 12$ are consistent with the calculated values of the Huang-Rhys parameter of ZnO in Refs. \cite{Janotti_PRL, Janotti_APL}. Importantly, the occupancy of the deep trap will be smaller than ca. 1$\%$ for small (or zero) values of the electron-phonon coupling, as can be appreciated in Figures 10-12.
\begin{figure}
\centering
\includegraphics[width=8cm]{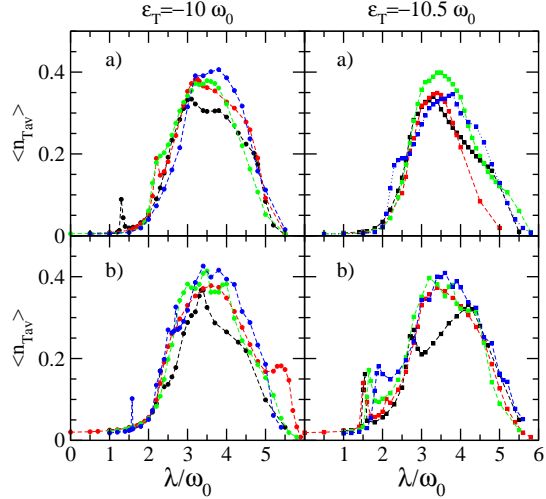}
\caption{Time-averaged mean trap occupancies $\langle n_{Tav} \rangle$  as a function of $\lambda$
for a) $V_0= \omega_0$ (top panels) and b) $V_0= 2\omega_0$ (bottom panels) and
for $\tilde \epsilon_T=-10 \omega_0$ (left panels) and 
$\tilde \epsilon_T=-10.5 \omega_0$ (right panels)
and the following values of $\gamma$:  $\gamma=0.3 \omega_0$ ($\Delta e=0.425 \omega_0$) (black symbols),
$\gamma=0.6 \omega_0$ ($\Delta e=0.83 \omega_0$) (red symbols), $\gamma=1.2 \omega_0$ ($\Delta e=1.7 \omega_0$) (green symbols) and 
$\gamma=2.4 \omega_0$ ($\Delta e=3.4 \omega_0$) (blue symbols).}
\label{fig12}
\end{figure}

\section{Conclusions}
In this work we have analyzed the effects of electron-phonon interaction in the dynamics of an electron that can be trapped to a localized state and detrapped to an extended band state in a simple model system. The system consists of a one-dimensional tight binding linear chain of a few sites, having a discrete set of energy levels mimicking the discrete levels of the conduction band of a small QD, that is connected at its end to another site, the trap, having a single energy level well below the conduction band, where the electron is allowed to interact with a local phonon of a single frequency.
In spite of its simplicity the time dependent model has no analytical solution but a numerically exact solution can be found at a relatively low cost.   
We start with the simplest possible system consisting of two sites since it can be approximated in different limits which allows us to identify the physical parameters controlling the time evolution of the electron-phonon system. The electronic motion is quasi-periodic in time, with oscillations around a mean value that are basic characteristics of the weak and strong coupling regimes of electron-phonon interaction and set the time scales of the system, ranging from 0.1 ps to 4 ps. However, the model yields a rich dynamics when hopping and electron-phonon interaction compete giving rise to a behavior reminiscent of blinking.  
The values of the time averaged trap occupancy strongly depend on the the strength of the electron-phonon interaction and can be as large as 40$\%$ when the coupling is most efficient, independently of other parameters. 
This situation is reached when the maximum of the trap density of states is near the bottom of the conduction band of the QD and the hopping parameter between the trap and the dot is larger than the phonon energy. 
One of the most interesting result of the present work is the formation of resonances. They are characterized by a trap occupancy that is a periodic function of time with large amplitude and period picturing an electron that is periodically trapped and detrapped. The resonances occur at specific values of the electron-phonon coupling parameter but only when several levels are allowed to coherently cooperate in the filling of the trap.
We have found them in both the two-sites and four-sites systems and for any value of the trap energy level.  Therefore we conclude that the formation of these resonances is a robust consequence of electron-phonon interaction in small, quantized, systems.  Electron-phonon interaction is an efficient mechanism that can provide ca. 50$\%$ filling of a deep trap state on a picoseconds or subpicoseconds time scale, much faster than radiative exciton decay occurring in time scales of tens of picoseconds to nanoseconds, while the occupancy of this state will be smaller than ca. 1$\%$ in the absence of electron-phonon coupling. In this work we have assumed that there is a single electron in the conduction band. Work in progress addresses the important issue of analyzing the effects of charging the QD by including more than one electron in the calculation, in which case an electron-electron Coulomb interaction should be considered at the trap site.
Further work includes the relaxation of the sudden switching-on of the hopping, since it might take some time for a trap to migrate to the surface, and the addition of hopping parameters to second neighbors. 

\section*{Acknowledgments}
We thank Alvaro Mart\'in-Rodero and Peter Apell for many useful discussions and a critical reading of the manuscript. Financial support from the Spanish
Ministry of Science and Innovation through the Mar\'ia de Maeztu Programme for Units of Excellence in R$\&$D
(CEX2018-000805-M) and the project RTI2018-099737-BI00 is acknowledged.

\end{document}